\documentclass[final,5p,times,twocolumn]{elsarticle}

 \usepackage{epsfig}
\usepackage{color}

\usepackage{amssymb}

\usepackage{graphicx} 
\usepackage{amsmath}
\usepackage{amsfonts}
\usepackage{amssymb}
\usepackage[applemac]{inputenc}
\usepackage[cyr]{aeguill}
\usepackage{textcomp}
\usepackage{subfigure}

\journal{Ultrasonics}

\begin{document}

\begin{frontmatter}



\title{A technique for measuring velocity and attenuation of ultrasound in liquid foams}


\author{J. Pierre$^1$, F. Elias$^2$, V. Leroy$^1$}

\address{1. Laboratoire MSC, Université Paris-Diderot, CNRS (UMR 7057), Paris, France\\ 2. Laboratoire MSC, Université Pierre et Marie Curie, CNRS (UMR 7057), Paris, France}

\begin{abstract}
We describe an experimental setup specifically designed for measuring the ultrasonic transmission through liquid foams, over a broad range of frequencies ($60$-$600\,$kHz). The question of determining the ultrasonic properties of the foam (density, phase velocity and attenuation) from the transmission measurements is addressed. An inversion method is proposed, tested on synthetic data, and applied to a liquid foam at different times during the coarsening. The ultrasonic velocity and attenuation are found to be very sensitive to the foam bubble sizes, suggesting that a spectroscopy technique could be developed for liquid foams. 

\end{abstract}


\end{frontmatter}

\newcommand{\ii}{\text{i}}
\newcommand{\ee}{\text{e}}
\newcommand{\real}{\text{Re}}
\newcommand{\imag}{\text{Im}}
\newcommand{\um}{\mu\text{m}}

\section{Introduction}
Acoustic spectroscopy~\cite{Dukhin2001,Challis2005} is a powerful tool to determine the particle size distribution in a dispersed media. It consists in measuring the effective attenuation and velocity of sound in the medium, over a broad range of frequencies (typically $1$-$100\,$MHz), and fitting those quantities with the appropriate model in order to obtain the particle size distribution. This technique gives real time measurements and is not destructive. It does not require dilution of the sample, and recent developments suggest that it could also give information on the local rheology of the medium~\cite{Bhosale2010}.

Acoustic spectroscopy is routinely used for dispersions of liquid droplets or solid particles, and commercial apparatuses are on the market for these kinds of media. Even gas bubbles can be inspected when their concentration is not too high~\cite{CBV2004,SLS2007}. But there is no spectroscopy technique available for liquid foams, even though those materials represent an important challenge in an increasing number of industrial areas (food, cosmetic or petroleum industry). Absence of ultrasonic spectroscopy can probably be explained by two difficulties that arise when dealing with the acoustics of liquid foams. First, the models used for other media are not directly applicable to liquid foams. Specific models have been developed~\cite{GOS1997,Kann2005}, but they still need to be experimentally validated. Which leads to the second problem: acoustic measurements in liquid foams are not straightforward. In particular, the measurements that exist in the literature~\cite{ZaK1991,OrS1993,MuF2002,SBB2006,Dau2011} do not cover a range of frequencies broad enough to measure a \emph{dispersion} of the acoustical properties, the key feature for extracting the size distribution. \\

In this article, we propose a new experimental setup for measuring the effective velocity and attenuation of ultrasound in liquid foams from $60$ to $600\,$kHz. 

\section{Experimental setup}
\subsection{Ultrasonic setup}
Our setup is based on the transmission of acoustic pulses through the sample to be characterized. As liquid foams are mainly composed of air, their acoustic impedance is low. Usual ultrasonic techniques that use piezoelectric transducers, and water as a coupling medium, face an impedance mismatch issue when dealing with such media. Air-borne acoustic waves thus seem more appropriate in terms of the amount of acoustic energy one can transmit through the sample. Moreover, the new generation of micromachined capacitance air transducers~\cite{GHB2001} does not suffer from the  bandwidth limitation that has made pulse techniques impossible for air-borne ultrasound. An additional benefit of working with air as coupling medium is to suppress the problem of water infiltration inside the sample, an issue particularly important for liquid foams.

\begin{figure}[h!]
	\centering
	\includegraphics[width =.8\linewidth]{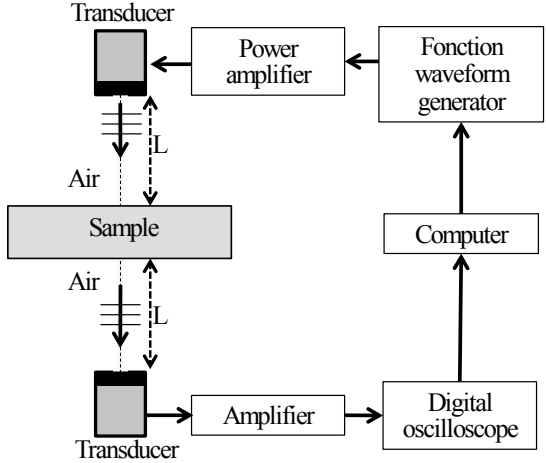}
	\caption{\textit{Scheme of the ultrasonic setup. A pair of broadband air-transducers measures the complex transmission through the sample from $60$ to $600\,$kHz.  }}
	\label{US_setup}
   \end{figure}

Figure~\ref{US_setup} shows a scheme of the ultrasonic setup. A function waveform generator, controlled by a computer, generated a gaussian pulse, amplified by a power amplifier (up to $150~$V for the most attenuating samples) and transmitted to the emitting air transducer (BAT-1, MicroAcoustic). The ultrasonic pulse then propagated through the sample before being received by an other transducer, pre-amplified ($40\,$dB) and recorded by a digital oscilloscope connected to the computer. Two gaussian pulses, centered around $150\,$kHz and $400\,$kHz, allowed us to explore the complex transmission from $60$ to $600\,$kHz. Note that we did not use the total bandwidth of the transducers because the attenuation was usually too high for frequencies higher than $600\,$kHz. Distance $L$ between the transducers and the cell was $9\,$cm, so that the cell was in the far-field of the transducers (the diameter of the active surface of the transducers was $1\,$cm).

\subsubsection{Design of the cell}
\begin{figure}[h!]
	\centering
	\includegraphics[width =7cm]{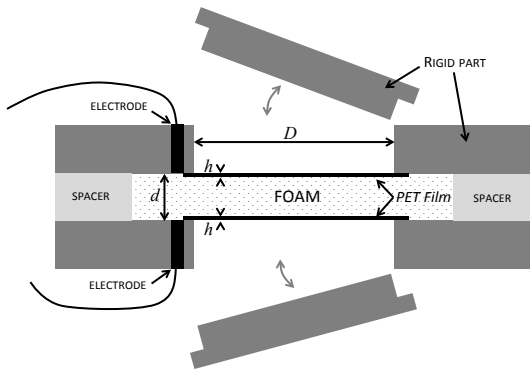}
	\caption{\textit{The liquid foam is contained in a thin-wall cell whose thickness $d$ is well defined. Two rigid plates are used to rigidify the walls when the cell is being filled. A pair of electrodes is also included in the cell for electrical conductivity measurements, which give access to the liquid content of the foam. }}
	\label{cell}
 \end{figure}

Liquid foams are not solid, which means one cannot cut a well-defined slab of foam and put it in the ultrasonic beam. On the other hand, using solid walls to maintain the foam pauses the problem of the impedance mismatch between the air and the walls. We designed a specific cell with very thin walls, depicted in figure~\ref{cell}. Circular holes with diameter $D=7\,$cm were made in two plastic plates ($18\times12\,$cm$^2$, $3\,$mm-thick) and covered with very thin PET (polyethylene terephtalate) films. The thickness of the films was chosen as small as possible ($h\simeq3\,\um$) in order to make them transparent to air-borne ultrasound.
 
Filling the cell was a critical task because thin films are flexible. Two solid plates that fitted exactly in the holes were clamped onto the walls to rigidify them during the filling process (see figure~\ref{cell}). Then, the liquid foam was poured onto one of the walls, two spacers with the desired thickness $d$ were placed on the edges, and the cell was closed with the second wall. The last step consisted in removing the rigid parts, making the cell ready for ultrasonic measurements. Note that the cell was placed horizontally to avoid a gradient of liquid fraction perpendicular to the ultrasonic beam. As liquid foams were found to be very attenuating for ultrasound, spacers were thin: $1.1 \pm 0.02\,$mm or even $0.5 \pm 0.02\,$mm for the most attenuating samples. However, it was checked that the diameters of the bubbles in the foam were smaller than $d$ (see section~\ref{bubsize}), ensuring a three-dimensional structure for the foam. Note that a benefit of the small thickness was that filling the cell did not require a large volume of foam (typically $8\,$mL were enough).


\subsubsection{Data processing}
\begin{figure}[th]
	\centering
	\includegraphics[width =0.9\linewidth]{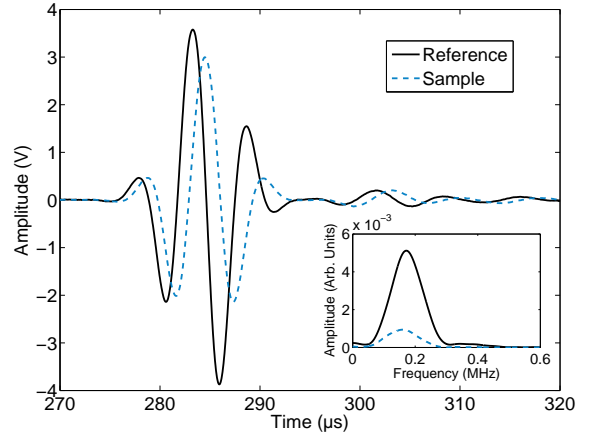}\label{Transmission_1PET_temporel}\\
	\caption{\textit{Measured signals through air (reference, solid line) and through one PET film (sample, dash line), for a gaussian pulse centered at $150\,$kHz. Inset: Fourier transforms of the two signals.}}
	\label{Pulse_1PET}
   \end{figure}

As an illustration of the data processing, we show the acquisition of the transmission through one PET film. Figure~\ref{Pulse_1PET} shows the two signals recorded by the oscilloscope without (Reference) and with (Sample) the film placed between the two transducers, for a gaussian pulse centered at $150\,$kHz. Note that even though the film is not perfectly transparent to ultrasound, a significant part of the energy is transmitted. The transit time is strongly affected: the pulses have a quadrature phase relationship. 
\begin{figure}[h!]
	\centering
	\subfigure[]{\includegraphics[width =0.75\linewidth]{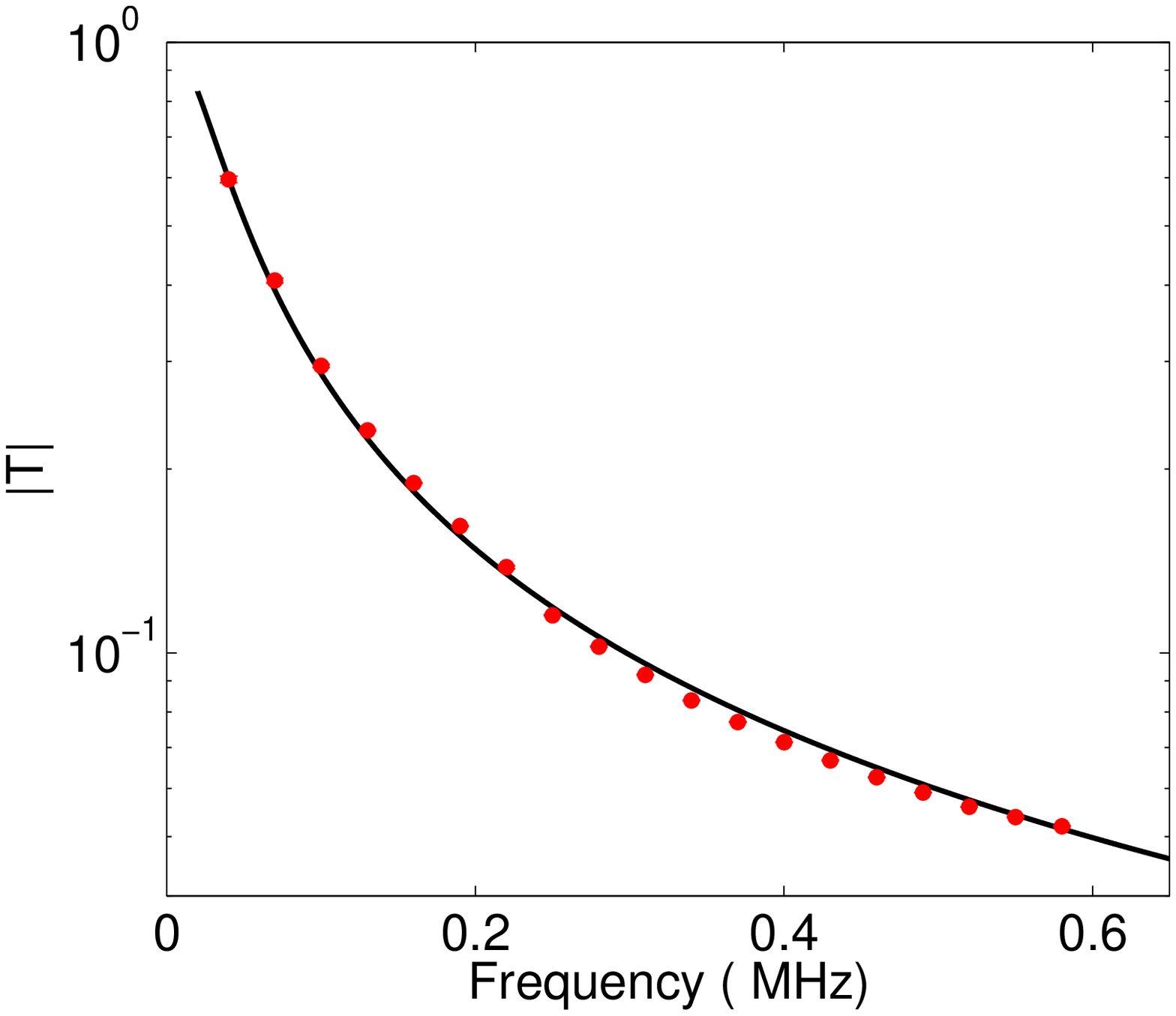}\label{Transmission_1PET_abs}}
	\subfigure[]{\includegraphics[width =0.75\linewidth]{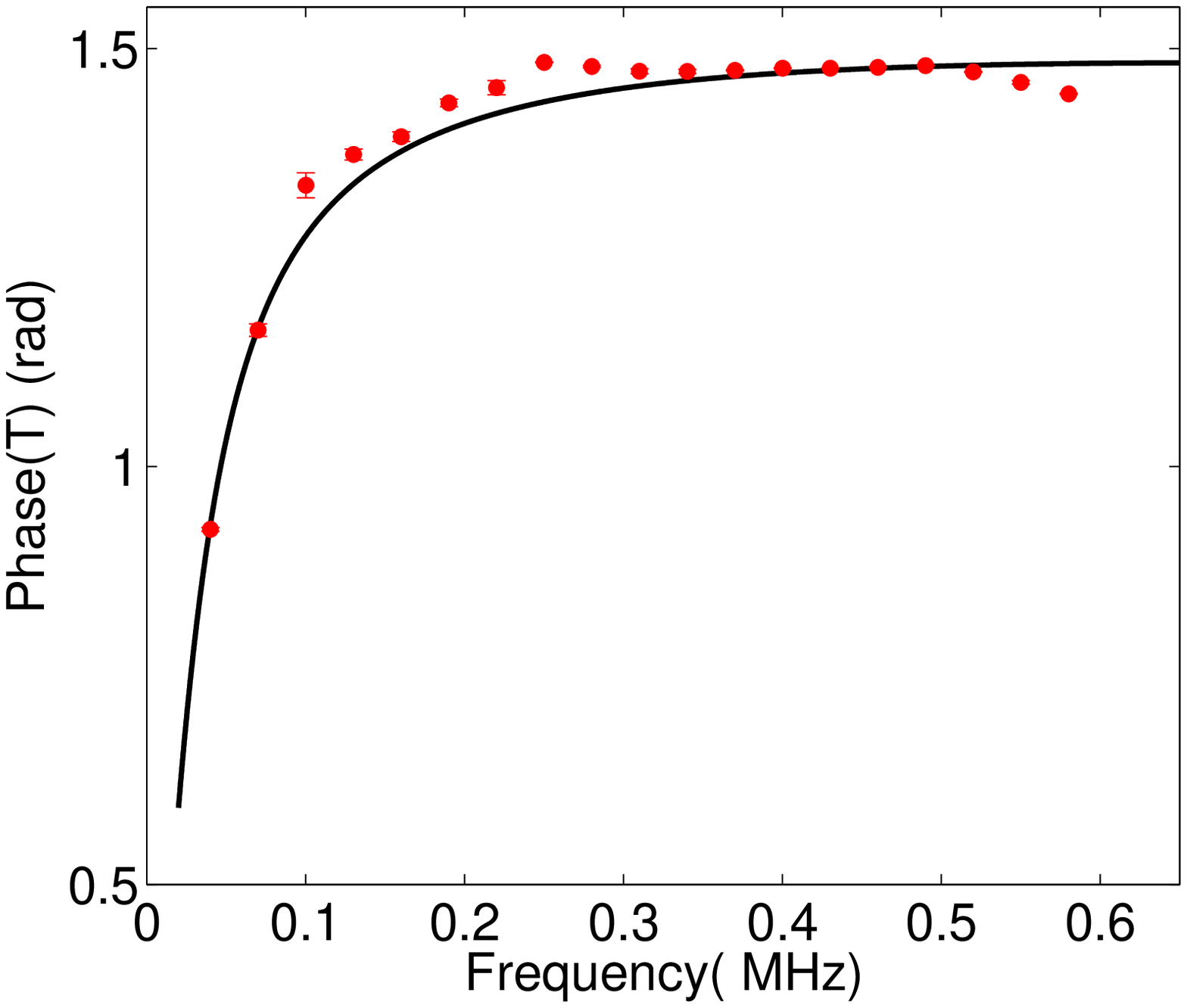}\label{Transmission_1PET_phase}}
	\caption{\textit{Transmission through one film. Amplitude (a) and phase (b) of the transmission coefficient are obtained by calculating the Fourier transforms of the signals shown in figure~\ref{Pulse_1PET}. Dashed lines correspond to the theoretical transmission through a film with thickness $h=3.1\,\um$.}}
	\label{Transmission_1PET}
   \end{figure}

The ratio between the Fourier transforms of the two temporal signals gives the complex transmission through the sample: $T=T_{SAM}/T_{REF}$, where $T_{REF}$ and $T_{SAM}$ are respectively the complex transmission without and with the sample.
Figures~\ref{Transmission_1PET}a and \ref{Transmission_1PET}b show the amplitude and the phase of $T$ as functions of the frequency. Note that the time limits of a pulse are not always clear. For instance, in figure~\ref{Pulse_1PET}, one can wonder if the small bump in the signal around $305\,\mu$s is part of the pulse, or a spurious signal. To solve this issue, we systematically made two different truncations: a short one that captured only the main pulse, and a broader one that included other small features. For the pulses of figure~\ref{Pulse_1PET},  the first truncation was $270$ to $300\,\mu$s, and the second one from $250$ to $320\,\mu$s. The differences between the two truncations appear as errobars in the amplitude and phase of the transmission (very small in figures~\ref{Transmission_1PET}a and \ref{Transmission_1PET}b because the signal over noise ratio was good).

\subsection{Foam production and characterization}
As the acoustic properties of liquid foams are known to strongly depend on their structure, we included other characterization techniques to the setup for measuring parameters such as the liquid fraction $\Phi$ of the foam and the bubble size distribution. Liquid foams have been the subject of an extensive literature, and many experimental techniques for characterizing them have been developed~\cite{Ca2010}. The technique used for the production of the liquid foam is known to have a great influence on the structure of the final product. We used a very simple technique known as the ``double-syringe'' technique. In one syringe, one takes a volume $V_\ell$ of liquid and completes with gas to a total volume $V$. Then the second syringe, empty, is connected to the first one and the mixture is pushed from one syringe to the other several times. With this simple process, the gas is sheared and divided into small bubbles. The size of the bubbles depend, among other parameters, on the size of the aperture that connects the two syringes. In our case, the syringes were connected with a standard female-female Luer Lock connector; the length and diameter of the cylindrical shearing zone were $16\,$mm and $4\,$mm, respectively. Note that the syringes should be rubber free, because rubber can limit the foamability of the solution.

For the foams presented in this article, the liquid phase was distilled water with $10\,$g/L of SDS (sodium dodecyl sulfate) and $0.5\,$g/L of xanthane. We used SDS as a surfactant because it gives a good foamability to the liquid, and xanthane because it increases the viscosity of the liquid phase, thus limiting the drainage of the foam~\cite{Sa2006}. The gaseous phase was air with traces of $C_6F_{14}$ (perfluorohexane) to decrease its solubility, which slows down the coarsening of the foam~\cite{Sa2006}.

\subsubsection{Liquid fraction measurement}

\begin{figure}[h!]
	\centering
	\includegraphics[width =0.7\linewidth]{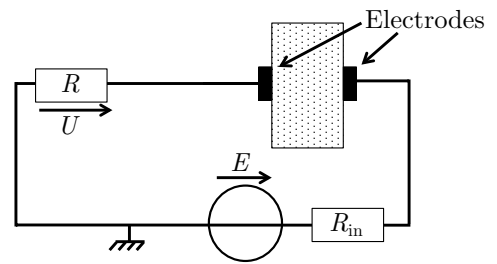}
	\caption{\textit{Electrical scheme of the conductivity setup. Measuring tension $U$ when the cell is filled with liquid and when it is filled with foam gives the relative conductivity of the foam, which is related to its liquid fraction.  }}
	\label{Cond_setup}
   \end{figure}
The advantage of the ``double-syringe'' technique is the relative control it offers on the liquid fraction of the foam, as one chooses the amount of liquid and gas taken to make the foam ($\Phi=V_\ell / V$). However, with standard small syringes and without specific precautions, the uncertainties on the volumes are such that the actual liquid fraction can noticeably differ from the expected one. We measured the liquid fraction by weighting the foam and found that, typically, when we aimed at a liquid fraction of $10\,\%$, the actual $\Phi$ was between $7$ and $13\,\%$. In order to have a precise and \textit{in-situ} determination of the liquid fraction we set up an electrical conductivity measurement~\cite{Le1978, FMS2005}. As shown in figure~\ref{cell}, two electrodes ($5\,$mm in diameter) were included in the walls of the cell. They were placed as close as possible from the edge of the film ($3\,$mm in practice), so that they probed the same part of the sample as the one probed by the ultrasonic beam. Figure~\ref{Cond_setup} presents the scheme of the electrical setup. The generator is set on an alternative tension of $1\,$kHz, with a voltage of $E=1\,$V and the resistance of the circuit is $R=1.2\,$k$\Omega$. The relative conductivity of the foam $\sigma_r$ is given by
\begin{equation}
\sigma_r=\frac{U_\ell}{U}\times \frac{E-U_\ell(1+R_\text{in}/R)}{E-U(1+R_\text{in}/R)}, \label{eqsigmar}
\end{equation}
where $R_\text{in}$ is the internal resistance of the generator, $U$ the tension measured on the resistor when the foam is in the cell, and $U_\ell$ the tension measured when the cell is filled with liquid.
From the relative conductivity, liquid fraction $\Phi$ is obtained thanks to the empirical law~\cite{FMS2005}:
\begin{equation}
\Phi=\dfrac{3\sigma_r(1+11\sigma_r)}{1+25\sigma_r+10\sigma_r ^2}.
\end{equation}
For the liquid foam we present in this article, we aimed at a liquid fraction of $5\,\%$ and obtained $\Phi=6\pm0.5\,\%$ according to the conductivity and weight measurements.

\subsubsection{Bubble size measurement}\label{bubsize}

\begin{figure}[h!]
	\centering
	\includegraphics[width =0.9\linewidth]{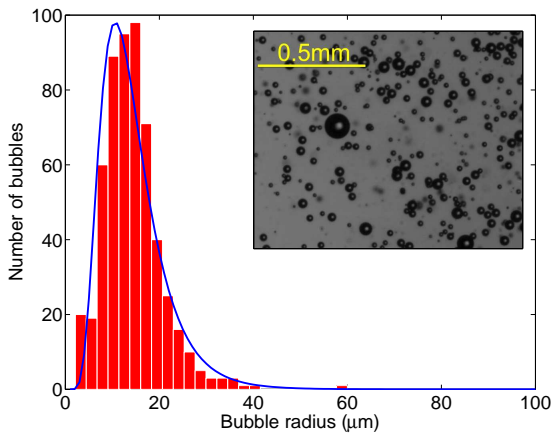}
	\caption{\textit{Image analysis of bubble rafts (see inset) gives the histogram of the bubble radii distribution. Here the distribution is fitted with a log-normal law (solid line); the median radius is $13\,\um$ and the polydispersity is $0.45$.  }}
	\label{Bubble_size}
   \end{figure}
Measuring the bubble size in a three-dimensional foam is very challenging. We adopted a quite intrusive method that consists in 
pouring a small volume of the foam onto the surface of the same liquid used for making the foam. One can then spread the bubbles in order to obtain a two-dimensional bubble raft, easy to image (see inset of figure~\ref{Bubble_size}). If it is assumed that the foam that remains in the syringe has the same time evolution than the one that has been put into the cell, this technique can even allow one to follow the evolution of the size distribution with time. Figure~\ref{Bubble_size} shows an histogram of the bubble radii, obtained for our example foam at $t=0$. It can be fitted by a log-normal law:
\begin{equation}
n(r)=\frac{n_{tot}}{r\epsilon \sqrt{2\pi}}\exp{\left( -\frac{(\ln(r/r_0))^2}{2\epsilon^2} \right)},
\end{equation}
where $n_{tot}$ is the total number of bubbles, $r_0$ the median radius of the bubble, and $\epsilon$ the log-normal standard deviation. Parameter $\epsilon$ is an indication of the polydispersity of the foam: $\epsilon=0.5$ means that there is a significant number of bubbles whose radius is half or twice the median radius.

\section{Transmission measurements}
\subsection{Ultrasonic characterization of the cell}
Being able to measure the acoustic properties of the foam contained in the cell requires a good knowledge of the cell itself. The first step is to measure the transmission through one film (see figure~\ref{Transmission_1PET}). As the impedance of PET is much larger than the one of air (see table~\ref{table}), and the thickness of the film is small compared to the wavelength in air ($\lambda\simeq0.5\,$mm at $600\,$kHz), the acoustic transmission through the film is well approximated by (see appendix~\ref{appendixA}):
\begin{equation}
T_\text{film}\approx \dfrac{ \text{e}^{ \text{i}(k_{p}-k_{a})h}}{1- \text{i} \pi \frac{f \rho_{p} h }{Z_a}},
\label{T_1FILM}
\end{equation}
where $f$ is the frequency, and $k$, $Z$, $\rho$ the wave numbers, acoustic impedances, and mass densities, respectively. Indices $a$ and $p$ refer to air and PET, respectively. As shown in figure~\ref{Transmission_1PET}, equation~(\ref{T_1FILM}) fits the experimental data very well for $h=3.1\,\um$.

\begin{table}[h!]
\begin{center}
\begin{tabular}{c c c c }
\hline
Medium & index & phase velocity & density \\
\hline
air & $a$ & $340\,$m/s & $1.2\,$kg/m$^{3}$ \\
PET & $p$ & $2540\,$m/s & $1400\,$kg/m$^{3}$ \\
\hline
\end{tabular}
\end{center}
\caption{Physical parameters for air and PET.  }
\label{table}
\end{table}

The amplitude of transmission through the film decreases as frequency increases: at $600\,$kHz, only $6\,\%$ of the amplitude is transmitted through the film. This is a limitation of our setup for high-frequency measurements. Note that the phase of the transmission also depends on frequency (see fig.~\ref{Transmission_1PET}b) and tends toward $\pi/2$ at high frequencies. It means that even though the film is very thin, $T_\text{film}$ is not purely real, which is an important feature for the inversion procedure (see section~\ref{inversion}).

\begin{figure}[h!]
	\centering
	\subfigure[]{\includegraphics[width =0.8\linewidth]{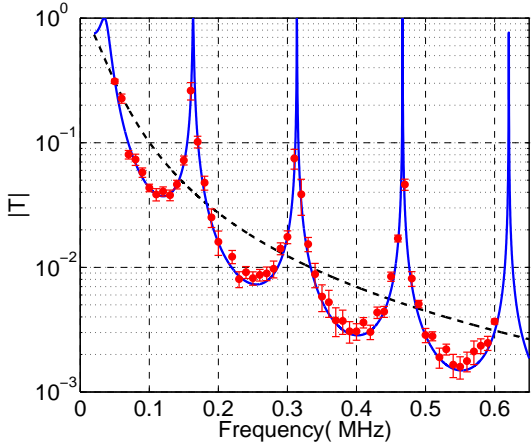}}
	\subfigure[]{\includegraphics[width =0.8\linewidth]{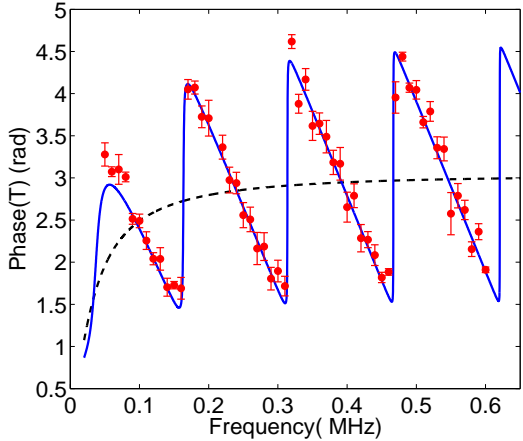}}
	\caption{\textit{Transmission through the empty cell. Amplitude (a) and phase (b) of the complex transmission though the empty cell are shown. The circles are data points, the continuous lines the theoretical prediction, and the dashed lines the transmission predicted through two films with no multiple reflections.}}
	\label{Transmission_2PET}
   \end{figure}

To make sure the roles of the two films were well understood, we also measured the transmission through the empty cell. Figure~\ref{Transmission_2PET} shows the amplitude and the phase of the transmission through this empty cell with a $d=1.1\,$mm spacer. Multiple reflections occur and give rise to very clear Fabry-Pérot resonances. The experimental data are successfully compared with the theoretical prediction $T_\text{cell}(k_a, \rho_a)$ (see appendix~\ref{appendixA}), which means that the cell is well characterized and ready for being filled with foam. Note that the first peak of transmission through the empty cell is particular. Indeed, the Fabry-Pérot criterion for the resonance to occur can be simply written as $k_a d=n \pi$ with integer $n\ge 1$, which gives frequencies of maxima for $f_n=n\times 155\,$kHz. It does not predict the first maximum which occurs at $37\,$kHz. This first maximum is actually a signature of the non purely real reflection coefficient of the films, which brings an additional phase shift and changes the condition of resonance.

\subsection{Transmission through liquid foams}
\begin{figure}[h!]
	\centering
	\subfigure[]{\includegraphics[width =.9\linewidth]{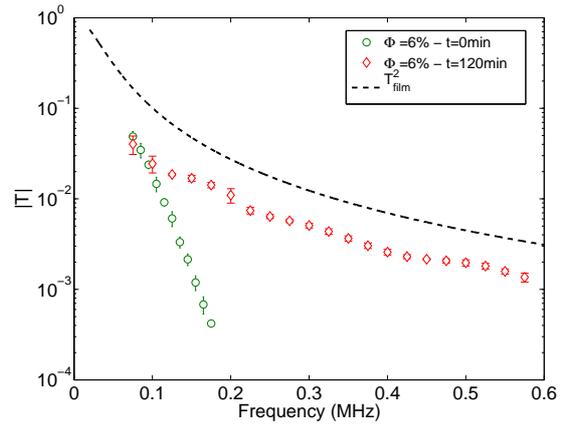}}
	\subfigure[]{\includegraphics[width =.9\linewidth]{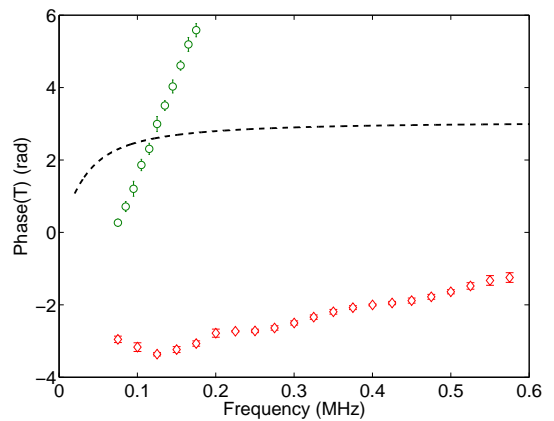}}
	\caption{\textit{Transmission through liquid foams. Amplitude (a) and phase (b) of the transmission through the cell filled with the same liquid foam at two different ages are shown. For comparison, the transmission predicted through two films with no multiple reflections is reported with dashed lines. }}
	\label{T_foam}
   \end{figure}

When the cell is filled with a liquid foam, the transmission does not show any Fabry-Pérot resonances (see figure~\ref{T_foam}), which is not a surprise given the high attenuation of ultrasound in liquid foams. We show results for a foam at two different ages, with a constant liquid fraction $\Phi=6\,\%$. Note that the attenuation was so high that a thin cell was used ($d=0.5\,$mm).  Even with this precaution, the signal could not be measured above $180\,$kHz at $t=0\,$min.

\section{Data analysis}\label{inversion}

\subsection{Principle}
The analysis of the experimental data is an inversion problem: from the measurement of the complex transmission through the cell, can we determine the velocity $c$ and the attenuation $\alpha$ of ultrasound in the medium? This is equivalent to determining the complex wavenumber $k=\omega/c + \ii \alpha/2$. As recalled in the appendix, the total transmission through the cell depends on the density and wavenumber in all the media involved, and on the thicknesses of the different layers. As the cell is well characterized, only $2$ quantities are unknown: $\rho$ and $k$, \emph{i.e.} the density and the wavenumber for the medium that fills the cell. If we note $T_\text{exp}$ the experimental transmission measured through the cell, the question is then to know whether we can find $\rho$ and $k$ such that $T_\text{cell}(k,\rho)=T_\text{exp}$.

A first simple approach is to consider that, as there is no multiple reflection in the liquid foam, $T_{cell}(k,\rho)$ reduces to $T_\text{film}^2\ee^{\ii (k-k_{a})d}$ because the acoustic wave just goes across the two films and the foam. This leads to 
\begin{equation}
k=k_{a}+\dfrac{1}{\ii d}\ln \left(\dfrac{T_{exp}}{{T_{film}}^2}\right).
\label{approxT}
\end{equation}

However, the transmission through the film is modified by the presence of the foam. The actual transmission through the cell is given by (see appendix~\ref{appendixA})
\begin{equation}
T_\text{cell}(k,\rho) = \frac{4Z_a \rho \omega/k}{(Z_a+\rho \omega/k)^2}\times \frac{\text{e}^{\ii (k-k_a) d}}{\left[1-\ii k_p h\left(\frac{Z_p}{Z_a+\rho \omega/k}\right)\right]^2}, \label{EqAinverser}
\end{equation}
which is much more challenging to inverse. First, both $\rho$ and $k$ are involved in the equation, which means that one of them needs to be determined by another measurement. Let us assume that we know $\rho$. Even with this simplification, $k$ cannot be extracted analytically from equation~(\ref{EqAinverser}). We used an iterative process: the first guess $k^{[0]}$ is given by eq.~(\ref{approxT}) and the iteration proceeds with
\begin{equation}
k^{[n+1]}=\frac{1}{\ii d}\ln \left(\frac{T_\text{exp}}{T_\text{cell}(k^{[n]},\rho)\text{e}^{-\ii k^{[n]}d}}\right).
\label{eqrec}
\end{equation}

Usually, $10$ iterations were enough to obtain a good convergence.

\subsection{Test on synthetic data}
We calculated $T_\text{cell}(k,\rho)$ for different $k$ and applied the inversion method to the synthetic data obtained. Convergence of the method was found to be excellent, providing that multiple reflections in the cell could be neglected (\emph{i.e.}, $\alpha d$ should be large enough). 

\begin{figure}[h!]
	\centering
	\subfigure[]{\includegraphics[width =.8\linewidth]{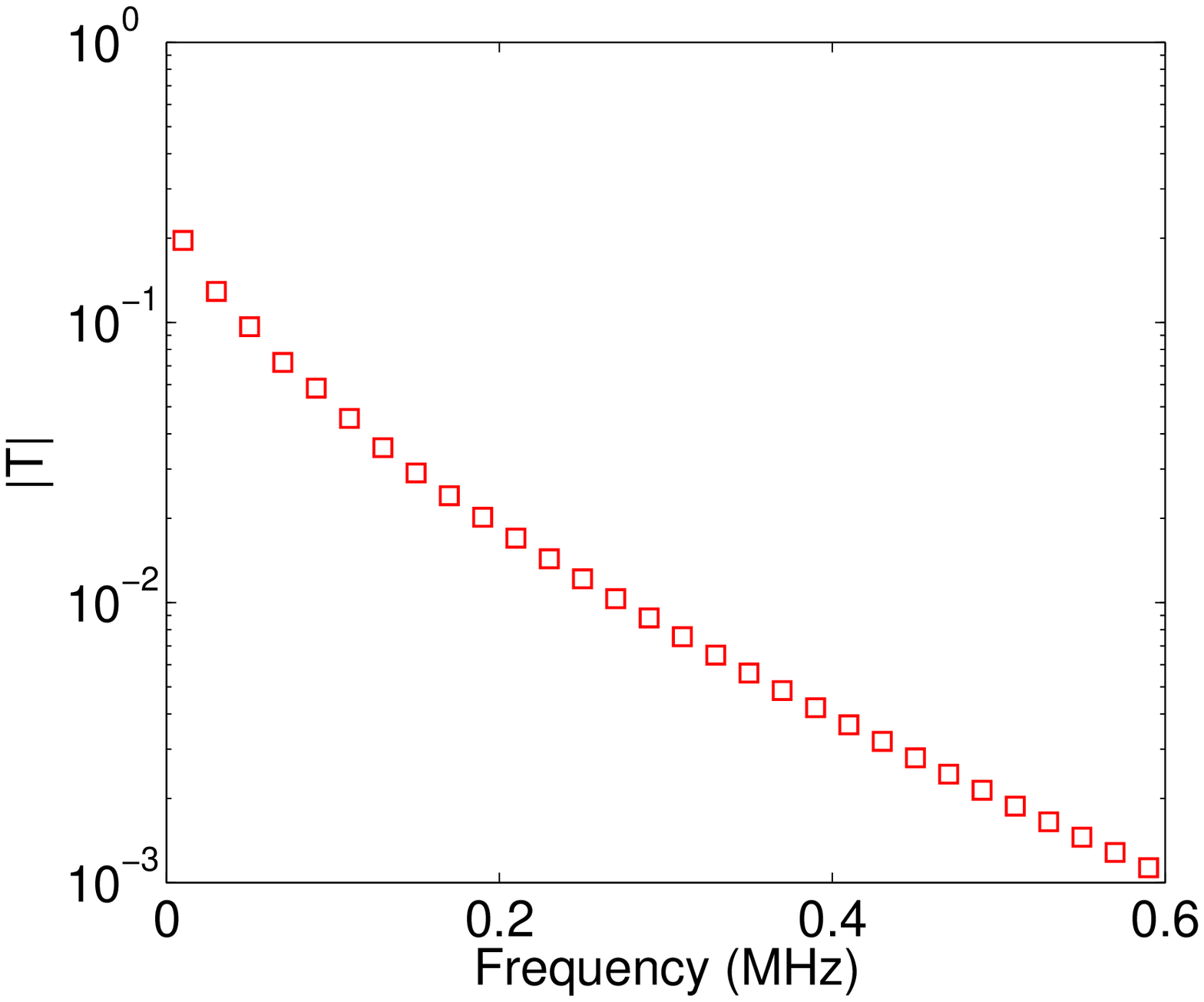}}
	\subfigure[]{\includegraphics[width =.8\linewidth]{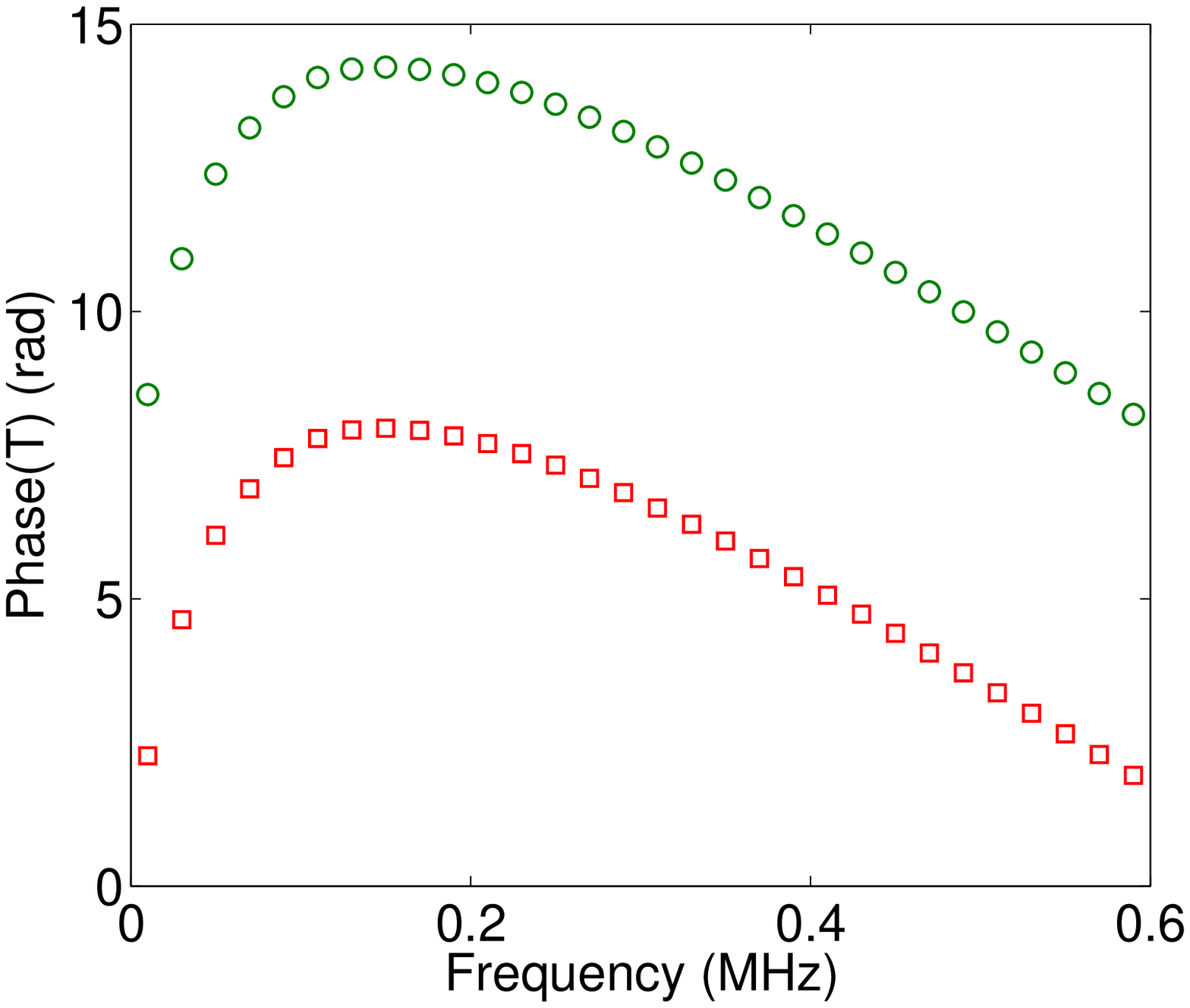}}
	\caption{\textit{Synthetic data are generated by calculating the amplitude (a) and phase (b) of $T_\text{cell}(k,\rho)$ for known values of $k$ and $\rho$. The phase is known modulo $2\pi$. As an example, the calculated phase (rectangles) can be affected by a $+2\pi$ shift (circles), resulting in a different result for the inversion (see text). }}
	\label{synthetic_data}
   \end{figure}

\begin{figure}[h!]
	\centering
	\subfigure[]{\includegraphics[width =.8\linewidth]{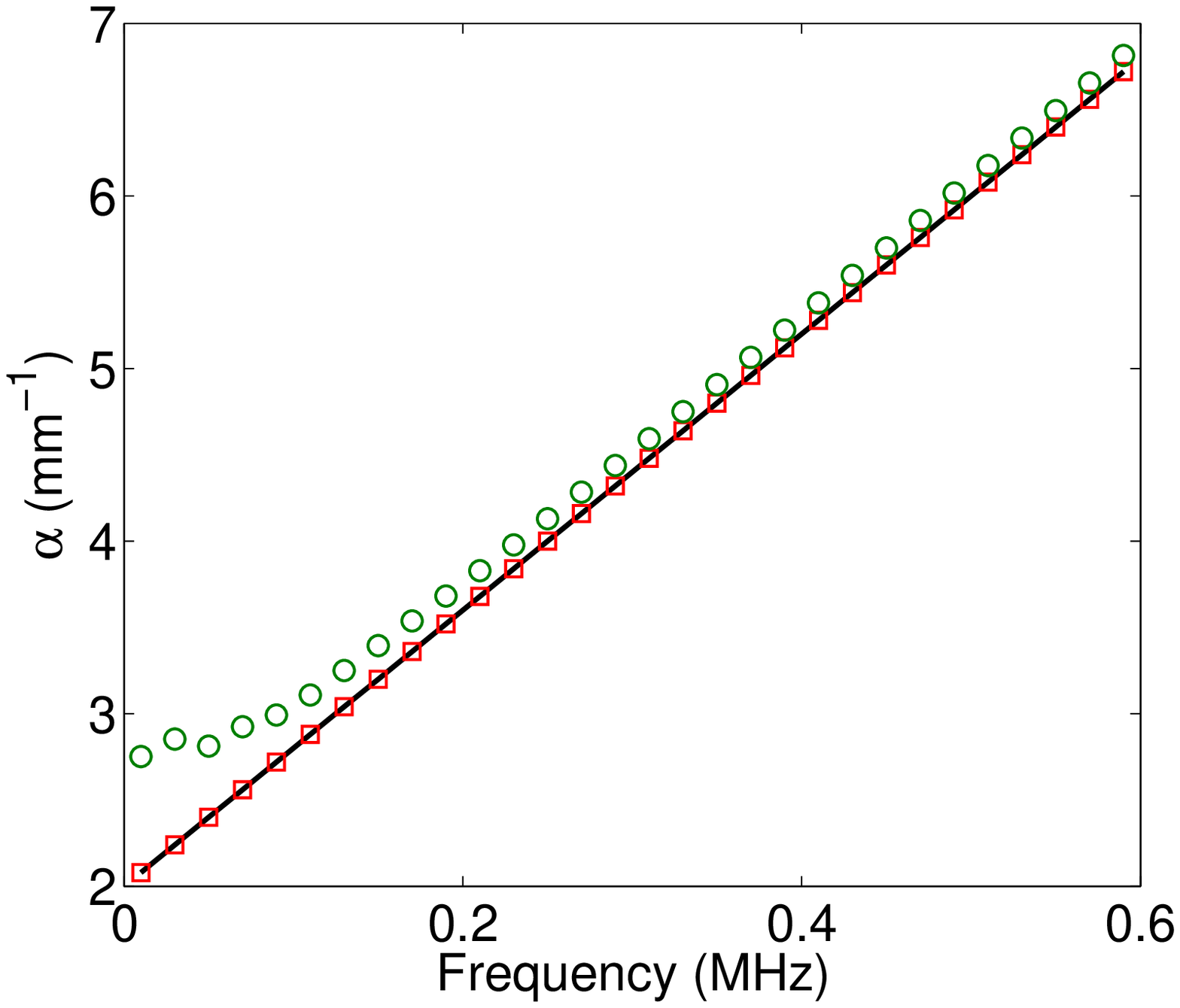}}
	\subfigure[]{\includegraphics[width =.8\linewidth]{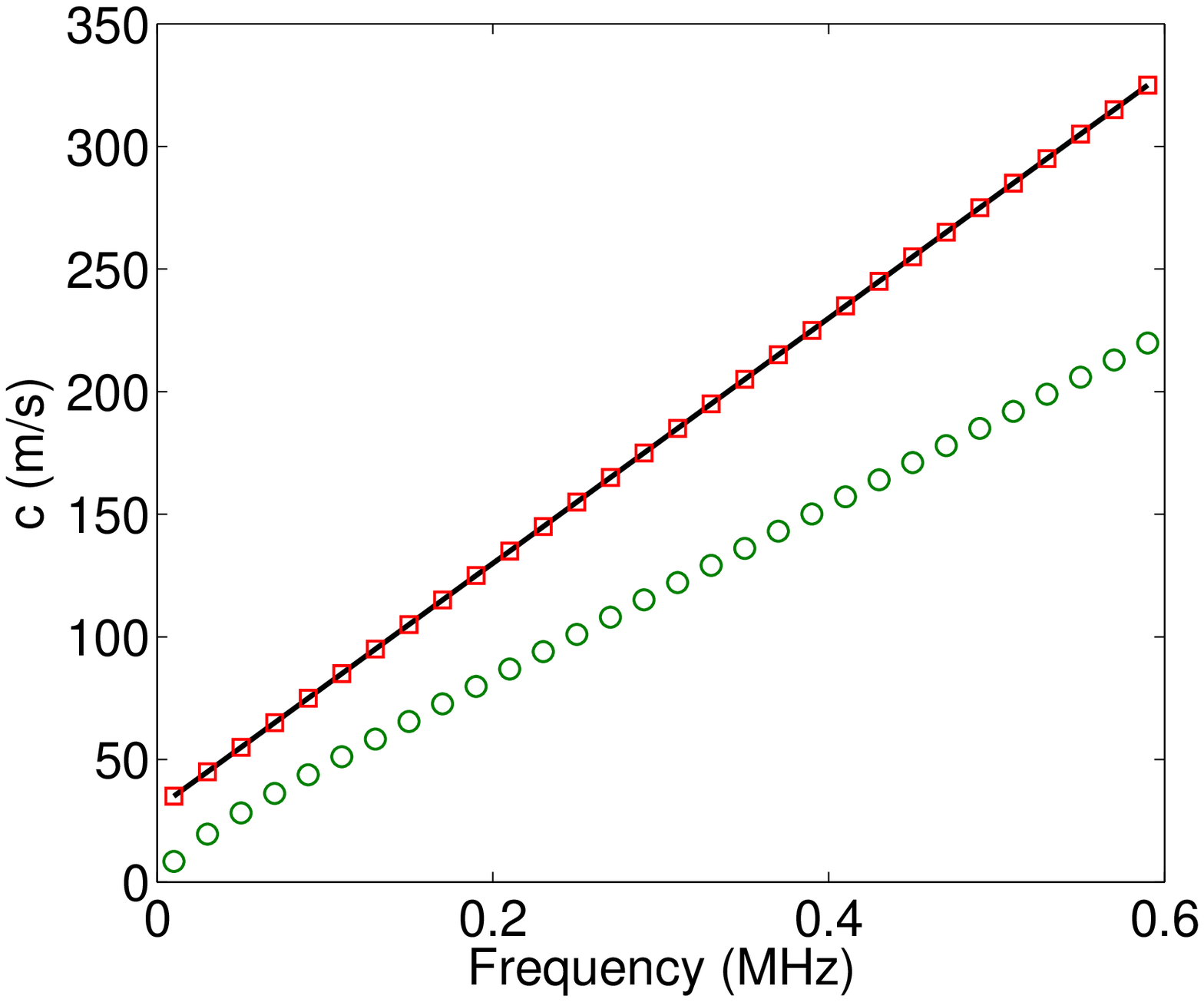}}
	\caption{\textit{When the inversion method is applied to the synthetic data of figure~\ref{synthetic_data}, one can find back the attenuation (a) and phase velocity (b) in the test medium. Different results are obtained for a zero phase shift (rectangles) or a $2\pi$ phase shift (circles). The true attenuation and velocity are shown in solid lines. }}
	\label{synthetic_data2}
   \end{figure}

Figure~\ref{synthetic_data} shows a typical result for a test medium with $\rho = 50\,$kg/m$^3$, $c=(30+500f)\,$m/s and $\alpha=(2+8f)\,$mm$^{-1}$, where frequency $f$ is in $MHz$. As a first step, we calculated the complex transmission through a fictive cell filled with the test medium (see appendix~\ref{appendixA}). Amplitude and phase of the complex transmission through the cell are shown in figure~\ref{synthetic_data}. A second step consisted in applying the inversion method to those synthetic data, which gave back the actual velocity and attenuation (figure~\ref{synthetic_data2}). 

Note that, as for any phase measurements, the phase of the transmission is measured modulo $2\pi$, which can affect the value found for the attenuation and the velocity. As an example, figure~\ref{synthetic_data2} shows how the inversion is affected when a global $2\pi$ phase shift is applied to the phase of the transmission. The attenuation is only slightly changed, but the velocity is significantly lowered. Determining the correct $2\pi$ shift for the phase has always been a problem for ultrasonic velocity measurements~\cite{PeP2003}. In many cases, the ambiguity is resolved because one knows that the velocity should be within a reasonable limit, or because incorrect $2\pi$ shifts lead to an unexpected dispersion. But for media such as liquid foams, many options give realistic results: both results in figure~\ref{synthetic_data2}b look plausible, for example. The solution to the problem is found by noting that the phase of the transmission should tend to zero for zero frequency~\cite{PeP2003}. Then, providing that broadband measurements are available, one can unambiguously determine the correct phase (see figure~\ref{synthetic_data}b for example). However, it requires the assumption that the phase has a smooth behavior for frequencies below the lowest experimental frequency.

\subsection{Experimental results for liquid foams}

As an illustration, we show that our technique allows us to follow the time evolution of the acoustical properties of a foam. 

Applying the inversion method to the real data of figure~\ref{T_foam} requires first to determine $\rho$. We took Wood's law~\cite{Wood1944} (also known as the mixture law), which assumes that the \emph{effective} mass density is the same as the \emph{actual} mass density: $\rho = \Phi \rho_\ell + (1-\Phi) \rho_g$, where $\rho_\ell$ and $\rho_g$ are the mass densities of the liquid and gaseous phases, respectively. For the liquid foam presented in this article, the inversion was thus performed considering that the mass density was $\rho = 60\,$kg/m$^3$. Note that other models exist and predict that the effective density can be complex and frequency dependent~\cite{ArA2007}.

\begin{figure}[h!]
	\centering
	\includegraphics[width =0.9\linewidth]{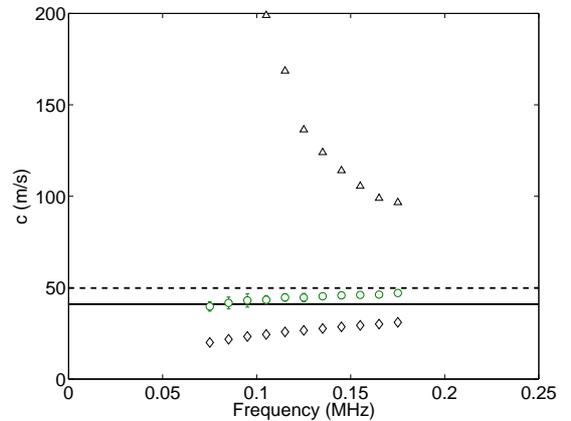}\label{alpha_foam}
	\caption{\textit{Phase velocity calculated from the experimental transmission of figure~\ref{T_foam}, using the inversion method for different ``2$\pi$'' shift: $0\pi$ (triangles), $+2\pi$ (circles) and $+4\pi$ (diamonds). }}
\label{2PI_shift}
  \end{figure}

As mentioned earlier, the phase is measured modulo $2\pi$.  As can be seen in figure~\ref{T_foam}b, the measured phase at $t=0\,$min needs a $+2\pi$ shift in order to follow our prescription of a ``zero phase at zero frequency''. Figure~\ref{2PI_shift} brings a confirmation that this is the correct phase correction. Indeed, the inversion was applied for $0$, $+2\pi$ and $+4\pi$ shifts. The $0\pi$ shift can be ruled out because it gives a very dispersive behavior. The $+4\pi$ shift gives a plausible result, but slightly more dispersive than the $+2\pi$ shift. Dispersion is not the only criterion: the value of the phase velocity found with the $+2\pi$ shift is in a close agreement with Wood's prediction: $c_\text{Wood}=\sqrt{100\kappa /\Phi/(1-\Phi)}$ for a gas at atmospheric pressure and a liquid with density of water, where $\kappa=1$ for isothermal transformations (solid line in Fig.~\ref{2PI_shift}) and $\kappa=1.4$ for adiabatic transformations (dashed line in Fig.~\ref{2PI_shift}). Other experiments at lower frequencies have found a velocity in liquid foams close to the isothermal Wood's velocity~\cite{OrS1993,SBB2006}.

\begin{figure}[h!]
	\centering
	\subfigure[]{\includegraphics[width =0.9\linewidth]{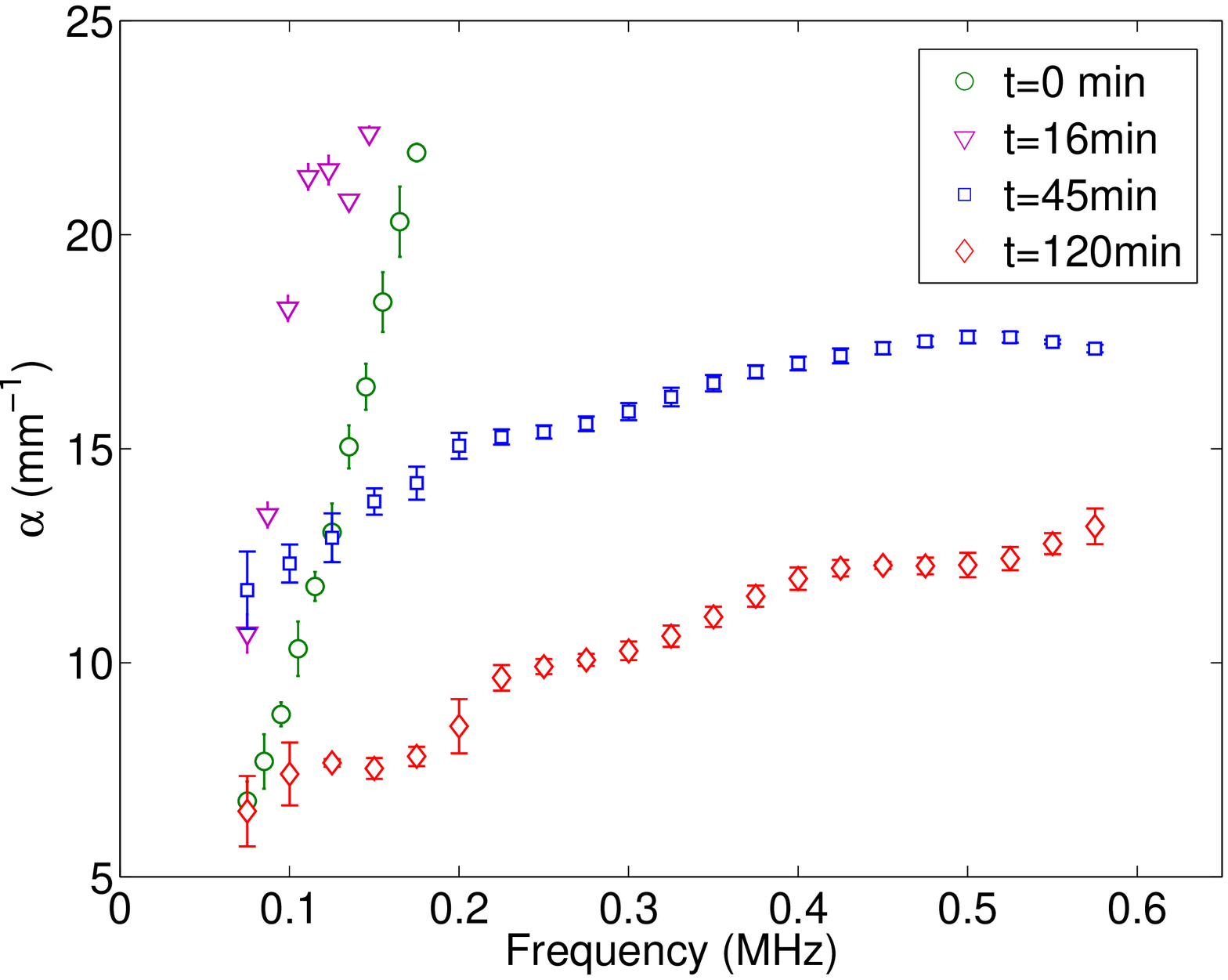}\label{alpha_foam}}
	\subfigure[]{\includegraphics[width =0.9\linewidth]{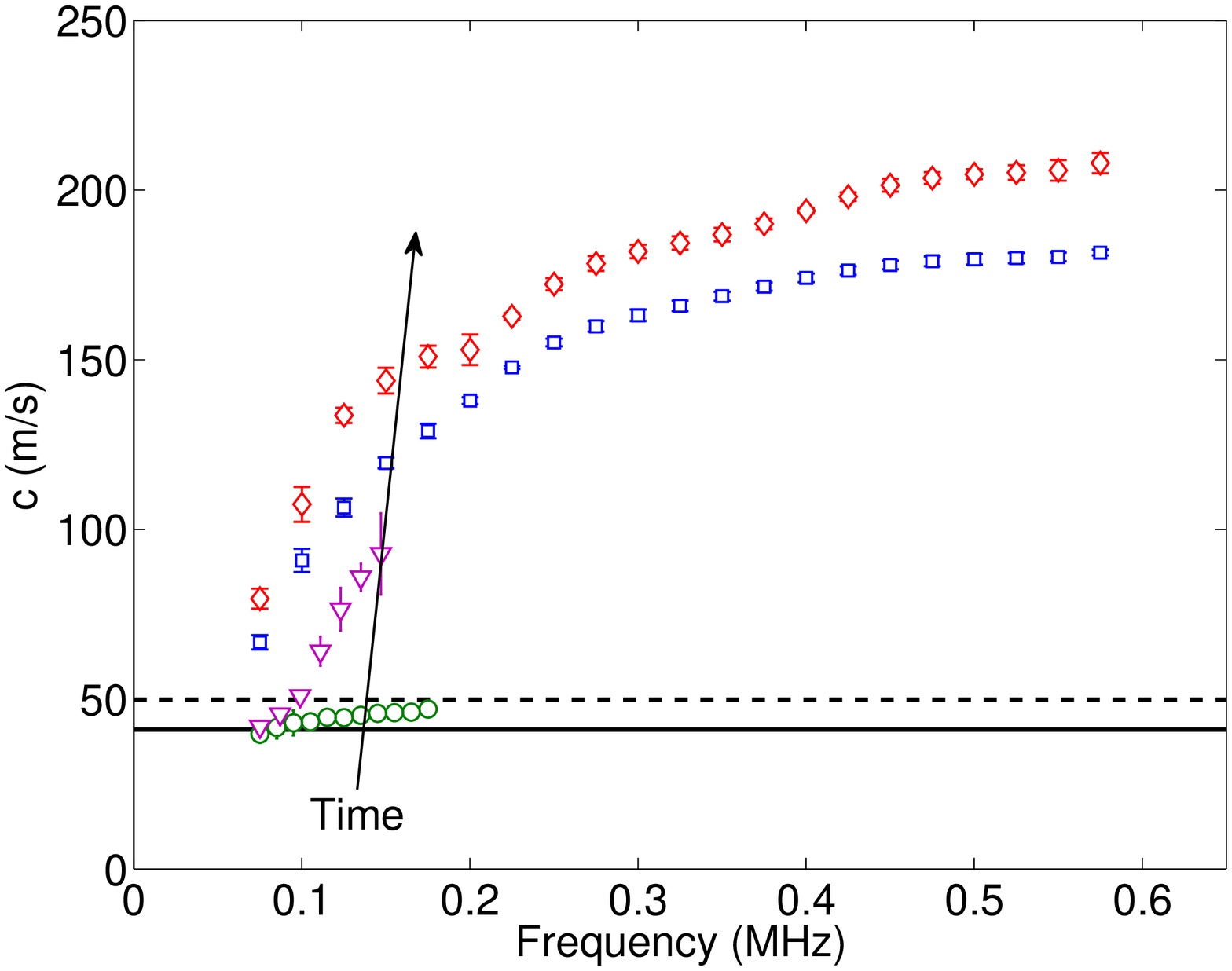}\label{v_foam}}
	\caption{\textit{Attenuation and phase velocity calculated from the experimental transmission of figure~\ref{T_foam}, using the inversion method. Attenuation and phase velocity for two intermediate ages of the foam are also reported.}}
\label{Results_foam}
  \end{figure}

Figure~\ref{Results_foam} shows the attenuation and phase velocity obtained for the liquid foam from $0$ to $120\,$min, \emph{i.e.} as the bubbles were growing from $13\,\mu$m (see Fig.~\ref{Bubble_size}) to $65\,\mu$m.  Errorbars were estimated from the three sources of uncertainty: liquid fraction ($\pm 0.5\,\%$), spacer thickness ($\pm 20\,\um$), and accuracy of the transmission measurement (see errorbars in Fig.~\ref{T_foam} for example). Even though the liquid content remains the same, the liquid foam shows very different ultrasonic behaviours as it ages. When the bubbles are small, attenuation is large and very dispersive, whereas the velocity is almost constant at about $40\,$m/s.  Interestingly, when the bubbles grow, the attenuation becomes lower and the velocity is found to be dispersive, reaching a plateau at about $200\,$m/s for $t=120\,$min.
This dispersive behavior has never been observed in liquid foams. It suggests that the size of the bubbles is an important parameter for the acoustic properties of liquid foams. 

\section{Conclusion}
We have described a new experimental setup for measuring the ultrasonic transmission through liquid foams over a large range of frequencies ($60$-$600\,$kHz). The ultrasonic setup was combined with conductivity measurements and optical observations that allowed us to characterize the liquid foam (liquid fraction and bubble size distribution). We have shown that, providing that the mass density of the foam was known, the velocity and attenuation of ultrasound in the foam could be deduced from transmission measurements. As an illustration of the technique, the acoustic properties of a foam at different ages were presented. The ultrasonic attenuation was found to be strongly dispersive and affected by coarsening. For the foam with small bubbles (at $t=0\,$min with median radius of $13\,\um$) the velocity was well predicted by Wood's approximation. On the other hand, for larger bubbles, the velocity was very dispersive. Thus, there is a clear signature of the bubble sizes in the acoustic measurements, which is a good indication that ultrasonic spectroscopy can be envisaged in liquid foams.
Note that an important feature of our experimental setup is real-time measurements (one acquisition is taken in less than $30\,$s). This is a major advantage for inspecting media such as liquid foams, whose time evolution is usually significant, or even crucial for some applications.

Further studies are needed for determining the exact link between the structure of the foam and its ultrasonic response. In particular, a systematic comparison of the experimental findings with the theoretical models should be undertaken. Hopefully, our setup might help for a better understanding of how ultrasound propagates in a liquid foam, an heterogeneous medium whose acoustic properties are still not clearly known.

\section*{Acknowledgement:} Support for the French Agence Nationale de la Recherche (project SAMOUSSE, ANR-11-BS09-001) is gratefully acknowledged. The authors thank Reine-Marie Guillermic and Julien Bonaventure for numerous discussions, as well as the GDR ``Mousses et Émulsions'' for its stimulating scientific environment.

\appendix
\section{Acoustic transmission through a three-layer system}\label{appendixA}
\begin{figure}[h!]
	\centering
	\includegraphics[width =5cm]{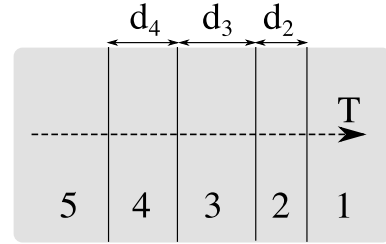}
	\caption{\textit{The transmission through $5$ successive layers is considered. Note that the layers are numbered backward compared to the direction of propagation.}}
	\label{fig5layer}
  \end{figure}

When an acoustic wave propagates through a layered media, such as the cell considered in this paper, many reflected waves are involved. For $5$ media (see figure~\ref{fig5layer}) the total transmission is given by~\cite{Bre1960}
\begin{eqnarray} 
T_{5\to1}
  &=& \frac{Z^{\text{\text{in}}}_4 +Z_4}{Z^{\text{in}}_4 +Z_5}\mathrm{e}^{\ii k_4 d_4}
  \times \frac{Z^{\text{in}}_3 +Z_3}{Z^{\text{in}}_3 +Z_4}\mathrm{e}^{\ii k_3 d_3} \nonumber\\
  && \times \frac{Z^{\text{in}}_2 +Z_2}{Z^{\text{in}}_2 +Z_3}\mathrm{e}^{\ii k_2 d_2}
  \times \frac{2Z_1}{Z_1 + Z_2}, 
  \label{eqbre}
\end{eqnarray}
where $d_i$ is the thickness of layer $i$, $Z_i=\rho_i
\omega/k_i$, $k_i$ stands for the impedance and the wave vector in
medium $i$ respectively, and $Z^\text{in}_i$ is the input
impedance for layer $i$, which is given by
\begin{subequations}
\begin{eqnarray}
    Z^{\text{in}}_1 &=& Z_1,\\
     Z^{\text{in}}_i &=& \frac{Z^{\text{in}}_{i-1}-\ii Z_i \mathrm{tan}(k_i d_i)}{Z_{i}-\ii Z^{\text{in}}_{i-1} \mathrm{tan}(k_i d_i)}Z_i \qquad \text{for}\quad i>1.
\end{eqnarray}
\label{eqZin}
\end{subequations}
With Eqs.~(\ref{eqbre}) and (\ref{eqZin}), one can predict the transmission through the cell filled with a medium whose density is $\rho$ and wavenumber $k$: $T_{cell}(\rho, k) = T_{1\to 5}/\text{exp}[{\ii k_a (d+2h)}]$, where $T_{1\to 5}$ is calculated with $k_1=k_5=k_a$, $\rho_1=\rho_5=\rho_a$, $k_2=k_4=k_p$, $\rho_2=\rho_4=\rho_p$, $d_2=d_4=h$, $d_3=d$ and $k_3=k$, $\rho_3=\rho$.

Several simplifications can occur. A first interesting situation is when layer $2$ is thin ($k_2d_2\ll 1$). Then the transmission from layer $3$ to layer $1$ simplifies into
\begin{equation}
T_{3\to 1} = \frac{2Z_1}{Z_1+Z_3}\times \frac{1}{1-\ii k_2 d_2 \left(\frac{Z_2^2+Z_1Z_3}{Z_2(Z_1+Z_3)}\right)}.
\end{equation}
If the acoustic impedance of the wall is large compared to the ones of the media it separates ($Z_2\gg Z_3$ and $Z_2\gg Z_1$), another simplification occurs:
\begin{equation}
T_{3\to 1} = \frac{2Z_1}{Z_1+Z_3}\times \frac{1}{1-\ii k_2 d_2 \left(\frac{Z_2}{Z_1+Z_3}\right)},
\end{equation}
which gives eq.~(\ref{T_1FILM}) if $Z_1=Z_3$.

Another interesting case is when the attenuation in medium $3$ is such that multiple reflections can be neglected (note that it also works if $d_3$ is large enough for multiple echos to be separated in the time domain), which leads to $Z^{\text{in}}_3= Z_3$. 
Thus, for a symmetrical configuration ($1=5$, $2=4$) with thin walls ($k_2d_2\ll1$) of high impedance ($Z_2\gg Z_3$ and $Z_2\gg Z_1$), if multiple reflection can be neglected in medium $3$, eq.~(\ref{eqbre}) reduces to
\begin{equation}
T_{5\to 1} = \frac{4Z_1Z_3}{(Z_1+Z_3)^2}\times \frac{\text{e}^{\ii k_3 d_3}}{\left[1-\ii k_2 d_2 \left(\frac{Z_2}{Z_1+Z_3}\right)\right]^2}.
\end{equation}


\bibliographystyle{model1-num-names}
\bibliography{<your-bib-database>}







 \end{document}